\pdfoutput=1 
\documentclass[fleqn,usenatbib]{mnras}

\usepackage{newtxtext,newtxmath}

\usepackage[T1]{fontenc}

\DeclareRobustCommand{\VAN}[3]{#2}
\let\VANthebibliography\thebibliography
\def\thebibliography{\DeclareRobustCommand{\VAN}[3]{##3}\VANthebibliography}

\usepackage{graphicx}	
\usepackage{amsmath}	
\usepackage{lineno}
\usepackage{xcolor}
\usepackage[normalem]{ulem}
\usepackage[separate-uncertainty=true]{siunitx}
\DeclareSIUnit\angstrom{\text {Å}}
\DeclareSIUnit\erg{\text {erg}}
\DeclareSIUnit\ph{\text {ph}}
\title[MAGIC detection of GRB 201216C at $z=1.1$]{MAGIC detection of GRB 201216C at $z=1.1$}

\author[H.~Abe~et.~al.]{\parbox{\textwidth}{\Large{
H.~Abe$^{1}$,
S.~Abe$^{1}$,
V.~A.~Acciari$^{2}$,
I.~Agudo$^{3}$,
T.~Aniello$^{4}$,
S.~Ansoldi$^{5,43}$,
L.~A.~Antonelli$^{4}$,
A.~Arbet Engels$^{6}$,
C.~Arcaro$^{7}$,
M.~Artero$^{8}$
% ...
}\\
{\small\textit{Affiliations are listed at the end of the paper}}
}
}

\date{}

\date{Accepted XXX. Received YYY; in original form ZZZ}

\pubyear{2023}

\begin{document}
\label{firstpage}
\pagerange{\pageref{firstpage}--\pageref{lastpage}}
\maketitle
\clearpage
\begin{abstract}
Gamma-ray bursts (GRBs) are explosive transient events occurring at cosmological distances, releasing a large amount of energy as electromagnetic radiation over several energy bands. We report the detection of the long GRB~201216C by the MAGIC telescopes. The source is located at $z=1.1$ and thus it is the farthest one detected at very high energies. The emission above \SI{70}{\GeV} of GRB~201216C is modelled together with multi-wavelength data within a synchrotron and synchrotron-self Compton (SSC) scenario. We find that SSC can explain the broadband data well from the optical to the very-high-energy band. For the late-time radio data, a different component is needed to account for the observed emission. Differently from previous GRBs detected in the very-high-energy range, the model for GRB~201216C strongly favors a wind-like medium. The model parameters have values similar to those found in past studies of the afterglows of GRBs detected up to GeV energies. 
\end{abstract}





\section{Introduction}
\label{sec:intro}

Gamma-ray bursts (GRBs) are sources exhibiting bright electromagnetic emission in two phases called \textit{prompt} and \textit{afterglow}. The former peaks at hard X-ray and soft gamma-ray energies, lasting between a fraction of a second and hundreds of seconds. In particular, the prompt temporal behavior shows short time scale variability down to milliseconds. Although its origin is not completely understood (for a review, see \citealt{kumarzhang}), recent evidence is pointing to a synchrotron origin \citep{zhao14,binbin16,oganesyan17,oganesyan18,oganesyan19}. The afterglow radiation partly overlaps with the prompt and evolves over longer timescales, up to several months after the GRB onset. The emission in this phase decays smoothly with time as a power law and it can be detected in several energy bands, from radio up to gamma rays, and is interpreted as synchrotron and Inverse Compton emission mostly from electrons accelerated in the external shock \citep{sari98,panaitescu00}.

GRBs are classified as short and long depending on whether their duration in terms of $T_{90}$, the time interval containing 90\% of the total photon counts, is shorter or longer than two seconds. While this observational definition is widely adopted, a more physical classification comes from the progenitor system at the origin of the bursts. In this context, short GRBs are thought to be produced as the result of the merger of binary systems of compact objects involving at least one neutron star (NS). The only confirmation of such association is the short GRB~170817A, which was detected in coincidence with a gravitational wave signal generated by a NS-NS merger \citep{PhysRevLett.119.161101,Goldstein_2017}. 
On the other hand, long GRBs are often associated with supernovae of type Ib/c, when detectable (e.g.\ if redshift is $z\lesssim1$). The supernova emission peaks several days after the GRB onset, when it outshines the decaying optical afterglow of the burst itself \citep{grb_sn_connection}.

The afterglow phase of GRBs has been studied in detail over several wavelength bands thanks to numerous instruments both ground-based (covering the radio and optical wavelengths, and VHE gamma rays) and space-based (detecting X-rays and gamma rays). Such observations have made it possible to trace the origin of the multi-wavelength afterglow emission to the synchrotron process \citep{afterglow_1,afterglow_2}. Such radiation is mostly produced by electrons accelerated at the so-called forward shock, when the GRB jet decelerates by interacting with the interstellar or circumstellar medium. Until recently, the afterglow was detected up to GeV energies by the \textit{Fermi}-LAT instrument, with some hints of a possible tail extending to higher energies \citep{2014Sci...343...42A}, 
where imaging atmospheric Cherenkov telescopes (IACTs) are more sensitive. The presence of emission in the very-high-energy (VHE, $E>100$\,GeV) range in the afterglow phase of GRBs was predicted, even before the operation of \textit{Fermi}-LAT and IACTs, in several theoretical models involving either leptonic or hadronic processes. A breakthrough was achieved in 2019, when the detection of VHE emission in the afterglow of three long GRBs was reported. The MAGIC collaboration first reported the detection of GRB~190114C \citep{grb190114c_gcn,grb190114c_discovery,grb190114c_mwl}, followed by GRB~180720B and GRB~190829A detected by the H.E.S.S. telescopes \citep{hess_grb180720b,hess_190829A}. The detection of such sources with IACTs confirmed the presence of an emission in the VHE range. In particular, the spectral and temporal analysis of GRB~190114C showed that such emission is associated with a component, separate from the synchrotron one, well explained by synchrotron-self Compton (SSC) radiation from electrons accelerated at the forward shock. A similar conclusion can be drawn for GRB~180720B (see e.g.\ \citealt{wang_grb180720b}), even though the multi-wavelength data available were not enough to perform a proper modeling. An unusual and controversial interpretation was put forward in the case of GRB~190829A. In \cite{hess_190829A} the authors suggested that the emission could be attributed to a single synchrotron component, which extends over nine orders of magnitude in energy up to the TeV domain. This requires an acceleration mechanism that is able to overcome the limit resulting in the so-called \textit{burnoff limit} for the energy of synchrotron photons (\citealt{dejager96,burnoff}).

The studies on this small sample of events shows how the understanding of the afterglow phase in the VHE range is far from complete. Currently only a few events have a detection at VHE (or evidence, as in GRB~160821B, see \citealt{grb160821b}), and different interpretations were proposed. However, the SSC scenario proved to be flexible and applicable to all the three GRBs detected at VHE. In order to investigate if such an interpretation may be universal to explain VHE afterglows, we present here the detection of the long GRB~201216C with the MAGIC telescopes. We use the available multi-wavelength data to model the broadband emission in the SSC scenario. We find that the SSC model provides a satisfactory interpretation of the MAGIC light curve and spectrum.

The paper is organized as follows. In Section~\ref{sec:observation}, we summarise all the observations available for GRB~201216C. In Section~\ref{sec:magic_data} we discuss the MAGIC observations and data analysis. The results are presented in Section~\ref{sec:results}. In Section~\ref{sec:mwl} we present the analysis of optical observations taken with the Liverpool Telescope and the other multi-wavelength observations that we use to model the emission with a synchrotron and SSC scenario (discussed in Section~\ref{sec:modeling}).
Finally, in Section~\ref{sec:conclusions} we summarise and discuss our findings.

\section{Observations of GRB~201216C}
\label{sec:observation}
GRB~201216C was detected by \textit{Swift}-BAT on December 16th 2020 at 23:07:31 UT (\citealt{grb201216c_swift_detection})\footnote{full GCN Circulars history at \url{https://gcn.gsfc.nasa.gov/other/201216C.gcn3}}, hereafter $T_0$. The burst was also detected by other space-based instruments including \textit{Fermi}-GBM, ASTROSAT  and Konus-Wind. The light curve by \textit{Swift}-BAT shows a multi-peaked structure\footnote{see \url{https://gcn.gsfc.nasa.gov/notices_s/1013243/BA/\#lc}} from  $T_0-\SI{16}{\second}$ to $T_0+\SI{64}{\second}$, with a main peak occurring at $\sim T_0+\SI{20}{\second}$.

GRB~201216C is classified as a long GRB, with a $T_{90}$ estimated to \SI{48(16)}{\second} in the $15-\SI{350}{\keV}$ band (\textit{Swift}-BAT, \cite{swift_gcn}) and \SI{29.953(572)}{\second} in the $50-\SI{300}{\keV}$ band (\textit{Fermi}-GBM, \citealt{GBM_GCN}).
The burst fluence in the $10-\SI{1000}{\keV}$ energy range between $T_0-\SI{0.003}{\second}$ to $T_0+\SI{49.665}{\second}$ is \SI{1.41(6)e-4}{\erg\per\cm\squared}. The 1-sec peak photon flux measured starting from $T_0+\SI{24.8}{\second}$ in the same energy band is \SI{54.9(6)}{\ph\per\second\per\cm\squared}.

Observations at different times by the VLT, FRAM-ORM and the Liverpool Telescope confirmed the presence of the optical afterglow. The position of the optical counterpart is consistent with the refined position provided by \textit{Swift}-XRT. 
VLT X-Shooter spectroscopy at $\sim T_0+2.4$\,hours, covering the wavelength range 3200-\SI{22000}{\angstrom}, allowed the measurement of the redshift, estimated\footnote{The value has been confirmed by the STARGATE collaboration via private communication.} to be $z=1.1$. Based on the VLT photometry, the steep photon index of optical data suggests a significant extinction, making GRB~201216C a dark GRB \citep{grb201216c_redshift}.

Assuming $z=1.1$, the isotropic energy release and peak luminosity of GRB~201216C in the $20-\SI{10000}{\keV}$ energy range are $E_{\rm iso,\gamma}=\SI{6.2(6)e53}{\erg}$ and $L_\textup{iso}=\SI{1.3(1)e53}{\erg\per\second}$.
With a rest frame spectral peak energy of $685\pm15$\,keV (\citealt{GBM_GCN}), GRB~201216C is consistent both with the Amati and Yonetoku correlations (\citealt{amati_relation, yonetoku_relation}).

The afterglow was also detected in the X-ray band by \textit{Swift}-XRT. The X-ray afterglow decay\footnote{see \url{https://www.swift.ac.uk/xrt_live_cat/01013243/}} can be described as a power law with temporal index $\alpha=\num{1.75(9)}$.

In the high energy range (\num{0.1}-\SI{1}{GeV}), \textit{Fermi}-LAT observed the burst from $T_0+\SI{3500}{\second}$ to $T_0+\SI{5500}{\second}$ but it did not detect any significant gamma-ray emission in such time interval, placing an energy flux upper limit of \SI{3e-10}{\erg\per\cm\squared\per\second} (95\% confidence level, $\SI{100}{\MeV}<E<\SI{1}{\GeV}$, see \citealt{grb201216c_fermilat}).

At higher energies, the burst was observed by HAWC starting at $T_0+\SI{100}{\second}$ up to $T_0+\SI{3600}{\second}$, resulting in a non significant detection \citep{hawc_gcn}.

Detection of radio emission was reported by \cite{rhodes22} from 5 to 56 days after the burst, from 1 to 10\,GHz. The radio flux at the time of detection is already decaying, although at a slow rate, except for the flux at 1\,GHz, for which the flux is increasing between 30 and 40 days.

Finally, the burst was observed by the MAGIC telescopes in the VHE range. Details of such observations are given in the following section.


\section{MAGIC observation and data analysis}
\label{sec:magic_data}

MAGIC is a stereoscopic system of two 17-m diameter IACTs situated at the Observatory Roque de los Muchachos (ORM), La Palma, Canary Islands. For short observations, as the ones usually performed for GRBs, the integral sensitivity achieved by MAGIC in \SI{20}{\minute} is about $20$\% of the Crab Nebula flux above \SI{105}{\GeV} for low zenith angles (see \citealt{aleksic_2016} for details on the telescopes performance).

MAGIC received the alert for GRB~201216C at 23:07:51 UT ($T_0+\SI{20}{\second}$) from the \textit{Swift}-BAT instrument. The MAGIC telescopes automatically reacted to the alert and, after a fast movement, they reached the target at 23:08:27 UT ($T_0+\SI{56}{\second}$). The observation was carried out in the so-called wobble mode around the coordinates provided by \textit{Swift}-BAT, RA:01h05m26s Dec:+16d32m12s (J2000). In local coordinates, the observation started at zenith \ang{37.1}, lasting up to 01:30:08 UT reaching zenith \ang{68.3}. The weather conditions were very good and stable during all the data taking with a median atmospheric transmission value at \SI{9}{\kilo\meter} a.g.l from LIDAR measurements of 0.96, with 1 being the transmission of a clear atmosphere (see \citealt{lidar_description,lidar_corrections} for a description of the LIDAR instrument and correction of VHE data). The observation was performed under dark conditions.

MAGIC continued the observation on the second night for \SI{4.1}{\hour} from $T_0+\SI{73.8}{\kilo\second}$. The observational conditions were optimal with an average transmission above 0.9 at \SI{9}{\kilo\meter} and dark conditions. The zenith angle changed from \ang{17.0} to \ang{46.3} with culmination at \ang{11.7}. The data on the second night was taken with the analog trigger system Sum-Trigger-II (described in \citealt{MAGIC_SUMT}), which was not available during the first night of data taking. Sum-Trigger-II improves the sensitivity of MAGIC in the low-energy range below $\sim\SI{100}{\GeV}$. In particular, the trigger efficiency, compared to the standard digital trigger, is two times larger for Sum-Trigger-II at \SI{40}{\GeV}.

The data analysis is performed using the standard MAGIC Reconstruction Software (MARS; \citealt{mars_2013}). In order to retain as many low energy events as possible, an algorithm \citep{mataju,geminga} where the calibration and the image cleaning are performed in an iterative procedure was adopted. This image cleaning was applied to the GRB data, gamma-ray Monte Carlo data, and to a data sample taken on sky regions without any gamma-ray emission (used for the training of the particle identification algorithm). Data analysis beyond this level is performed following the prescriptions described in \cite{aleksic_2016}. The usage of Sum-Trigger-II, combined with the optimized cleaning algorithm, allows for a collection area an order of magnitude larger around \SI{20}{\GeV} when compared with the one obtained with the standard digital trigger.


\section{Results from the very-high-energy data}\label{sec:results}

In this section we show the results of the analysis performed on the data collected by MAGIC on GRB~201216C.

\subsection{Detection and sky map}
\label{subsec:detection}

\begin{figure}
\begin{center}
\includegraphics[width=1.0\columnwidth]{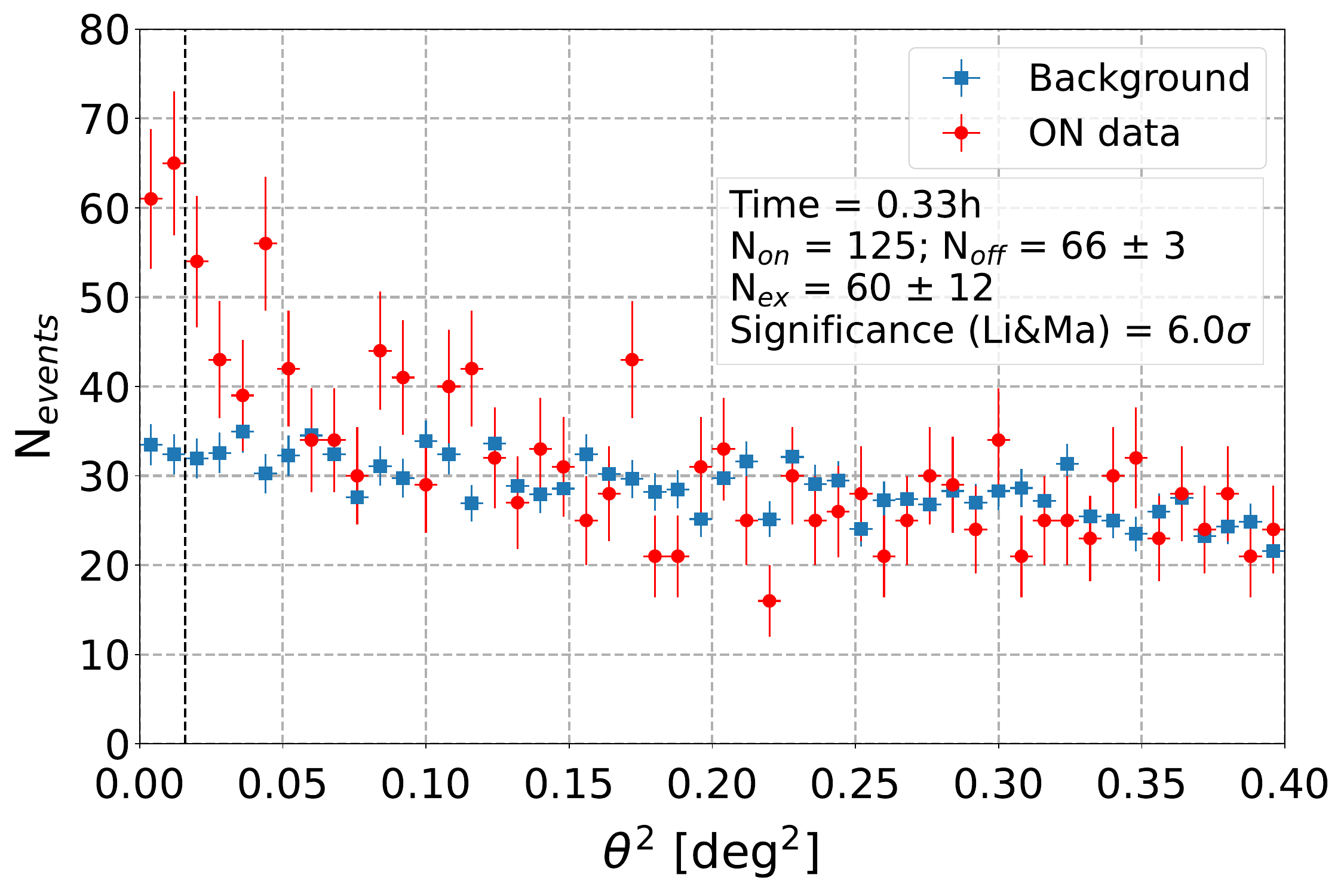}
\caption{$\theta^2$ distribution for the first 20 minutes of observation, see the main text for the definition. Both GRB (red circles) and background events (blue squares) are shown. The vertical dashed black line shows the value of the cut in $\theta^2$ used for the calculation of the significance.}
\label{fig:theta2}
\end{center}
\end{figure}

Fig.~\ref{fig:theta2} shows the distribution of the squared angular distance, $\theta^2$, for the GRB and background events (red circles and blue squares, respectively) for the first 20 minutes of data (from $T_0+\SI{56}{\second}$ to $T_0+\SI{1224}{\second}$). The significance of the VHE gamma-ray signal from GRB 201216C is 6.0\,$\sigma$, following the prescription of \cite{lima_1983}, confirming the significant detection of the GRB. For the computation of the significance, we apply cuts on $\theta^2$ and hadronness. The former is the squared angular distance between the reconstructed direction of the events and the nominal position of the source, taken from \textit{Swift}-BAT for GRB~201216C. The latter is a parameter which discriminates between gamma-like and background-like events, with gamma rays having hadronness values
close to zero. The cuts on $\theta^2$ and hadronness were optimized for a source with an intrinsic power-law spectrum with index $\alpha=-2$, later corrected considering the absorption by the extragalactic background light (EBL) according to the model by \cite{Dominguez2011}, hereafter D11. For the signal significance evaluation, the intrinsic spectral index for the cut optimization was chosen to be similar to the one found in the other GRBs detected at VHE, so without any prior knowledge of the actual value for this specific GRB (see Section~\ref{subsec:spectrum}). The corresponding energy threshold of the optimized cuts is 80 GeV defined by the peak of the energy distribution of the surviving simulated events.

\begin{figure}
\begin{center}
\includegraphics[width=0.75\columnwidth]{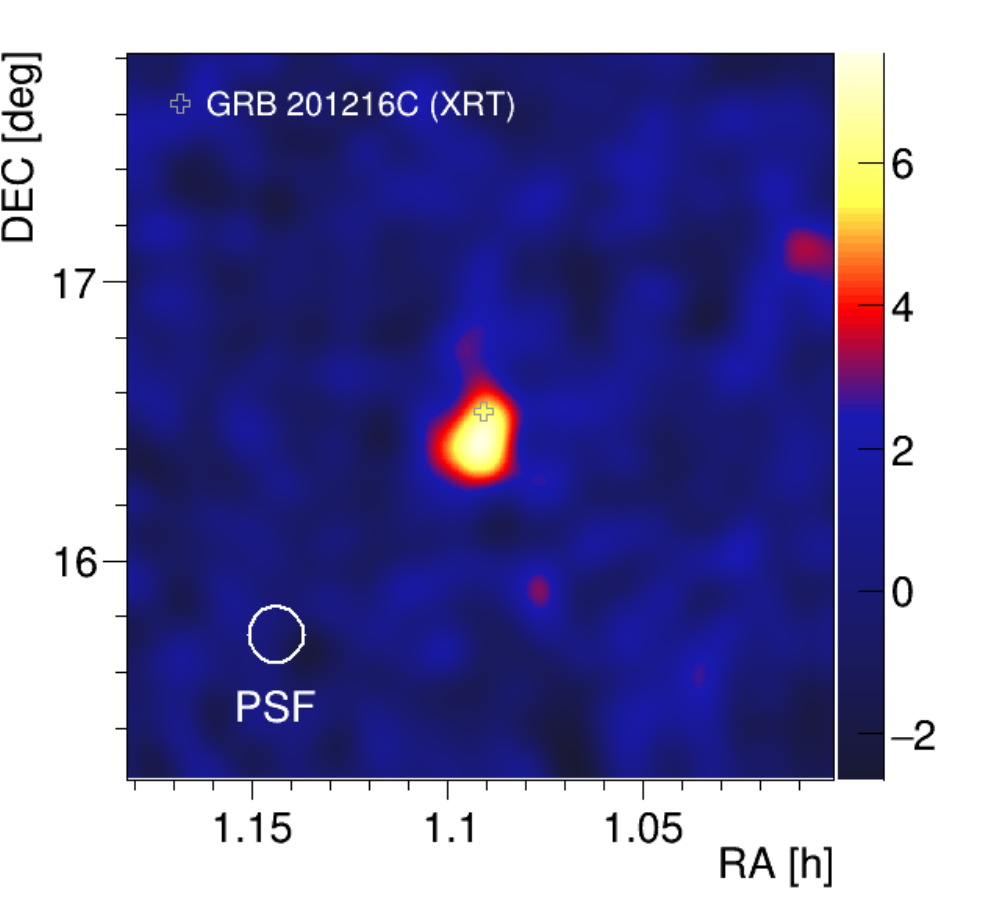}
\caption{Test-statistics sky map for the first 20 minutes of observation. The cross marker shows the position of GRB~201216C reported by \textit{Swift}-XRT. The white circle shows the MAGIC point spread function corresponding to 68\% containment.}
\label{fig:skymap}
\end{center}
\end{figure}

Fig.~\ref{fig:skymap} shows the test-statistics map in sky coordinates for the first 20 minutes of data. The same event cuts as for Fig.~\ref{fig:theta2} are used. Our test statistic is \cite{lima_1983} equation 17, applied on a smoothed and modeled background estimation. Its null hypothesis distribution mostly resembles a Gaussian function, but in general can have a somewhat different shape or width. In the sky map, the peak position around the center is consistent with the one reported by \textit{Swift}-XRT within the statistical error. The peak significance is above 6\,$\sigma$, which corroborates the detection.

\subsection{Average spectrum}
\label{subsec:spectrum}

\begin{figure}
\begin{center}
\includegraphics[width=0.9\columnwidth]{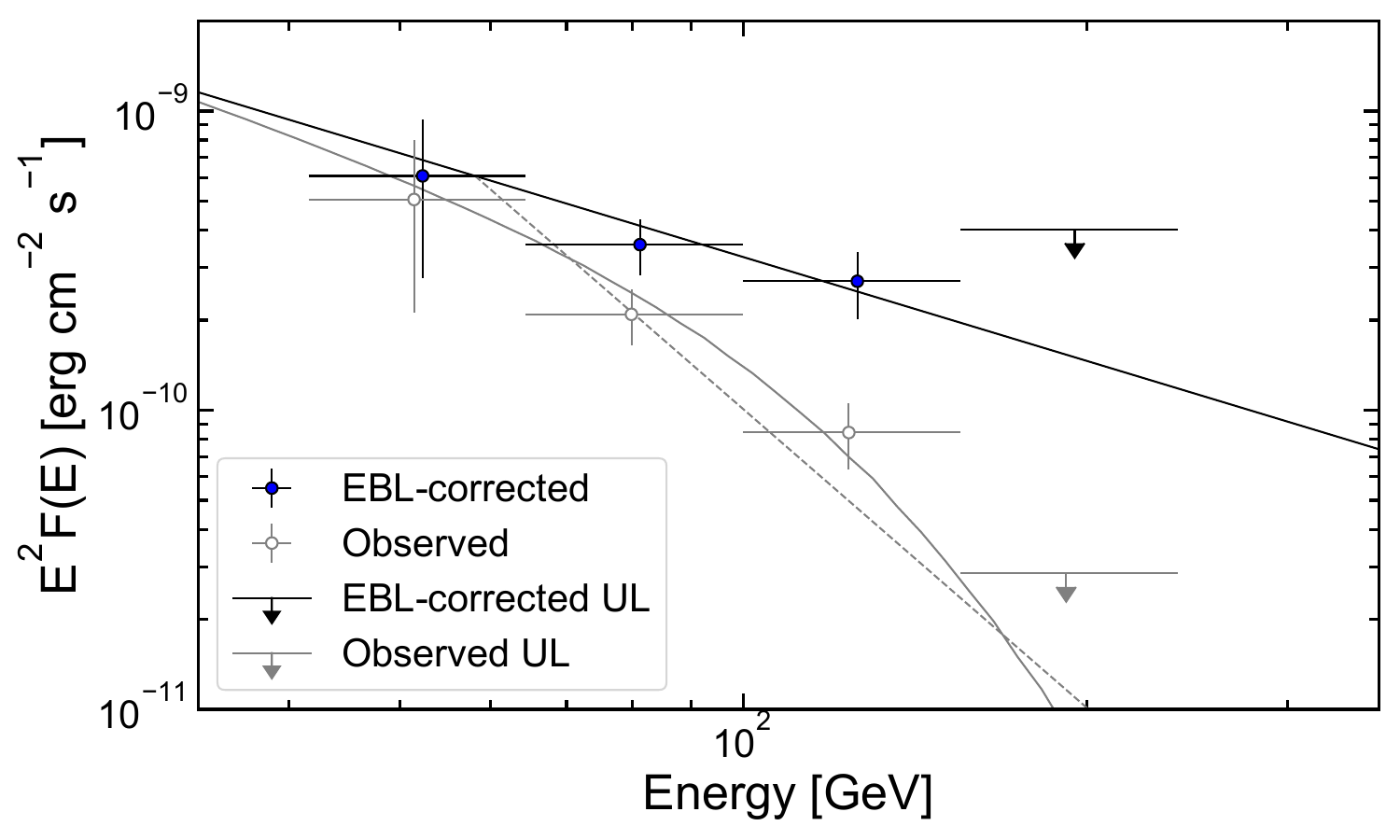}
\caption{Observed and EBL-corrected spectra for GRB~201216C as measured by the MAGIC telescopes during the first 20 minutes of observations, denoted by white and blue filled points respectively. The highest energy bin is a $2\sigma$ upper limit in each spectrum. The solid black and dashed grey lines represent the forward folding fits to the data points. The solid grey line is obtained from the intrinsic spectrum fit (black solid line) after the absorption by the EBL is taken into account, using the D11 model.}
\label{fig:time_integrated_spectrum}
\end{center}
\end{figure}

The average spectrum for the first 20 minutes of observation is shown in Fig. \ref{fig:time_integrated_spectrum}. The data points are the result of an unfolding procedure following the prescription of the Bertero method described in \cite{MAGIC_unfolding}. 
The best fit to the points is instead provided by the forward folding method (\citealt{forward_folding}). For the event cuts optimization, the adopted spectrum is an intrinsic power-law spectrum with an index $\alpha=-3$, which is close to the final estimated value (see below), later attenuated by EBL assuming the model D11 and $z=1.1$.
Because of the strong EBL absorption, the observed spectrum has a steep power-law index of $-5.32\pm 0.53$ (stat. only) above 50 GeV. The intrinsic (EBL-corrected) spectrum is consistent with a simple power-law function and shows a harder index of $-3.15\pm0.70$ (stat. only). The normalization factor at 100 GeV is \SI{2.03\pm0.39e-8}{\per\TeV\per\cm\squared\per\second} (stat. only). The highest energy bin around 200 GeV is a $2\sigma$ upper limit due to a large relative flux error about 100\%. 

The obtained spectrum suffers from systematic uncertainties coming from different sources. For such a steep observed spectrum, the uncertainty of the energy scale significantly affects the computed fluxes. We estimated the flux variation by shifting the light scale in the simulations during the forward folding procedure assuming the EBL model D11. We adopted a $\pm15$\% shift as prescribed in \cite{aleksic_2016}. The results are shown in Table \ref{tab:Escale_EBL_systematics}. When the energy scale is shifted by -15\%, the observed spectrum is shifted to the low-energy side resulting in a lower flux. The spectral index of the intrinsic spectrum is softened due to the smaller attenuation by EBL at lower energies. In case of the +15\% shift, the flux and the spectral index are shifted in the opposite direction. The obtained power-law index ranges from -3.19 in the -15\% case to -2.17 in the +15\% case. The normalization factor instead varies by a factor of 3. The spectral uncertainty originating from the energy scale is therefore significantly larger than the statistical errors.

\begin{table}
\begin{center}
\begin{tabular}{c|c|c|c}
    light scale & EBL & normalization [\SI{}{\per\TeV\per\cm\squared\per\second}] & index \\
    \hline\hline
    nominal & D11 & \SI{2.03\pm0.39e-8}{} & $-3.15\pm0.70$ \\
    \hline
    -15\% & D11 & \SI{1.14\pm0.25e-8}{} & $-3.19\pm0.52$\\
    \hline
    +15\% & D11 & \SI{2.99\pm0.53e-8}{} & $-2.17\pm0.57$\\
    \hline
    nominal & F08 & \SI{1.95\pm0.38e-8}{} & $-3.19\pm0.70$ \\
    \hline
    nominal & FI10 & \SI{2.76\pm0.54e-8}{} & $-2.65\pm0.73$ \\
    \hline
    nominal & G12 & \SI{3.99\pm0.77e-8}{} & $-2.45\pm0.71$ \\
\end{tabular}
\caption{Fitted power-law spectral parameters of the 20-minute average spectrum using different scales of the Cherenkov light amount and different EBL models. The tested EBL models are D11, F08, FI10, and G12 with the nominal light scale. The tested light scales are nominal, -15\%, +15\% with the D11 EBL model. The normalization energy is fixed to 100 GeV. The errors are statistical only. The resulting systematic errors are reported in the main text.}
\label{tab:Escale_EBL_systematics}
\end{center}
\end{table}

The VHE flux of GRB 201216C is also affected by the choice between available EBL models. At such high redhift $z=1.1$, EBL models show large differences in predicted attenuation factors. We compared the spectra calculated with four EBL models including D11 with the same unfolding method as the one used for Fig. \ref{fig:time_integrated_spectrum}. The three models besides D11 are \cite{Franceschini2008}, \cite{Finke2010}, and \cite{Gilmore2012} (hereafter F08, FI10, and G12, respectively). The results are shown in Table \ref{tab:Escale_EBL_systematics}. The power-law index ranges from -3.19 in the F08 case to -2.45 in the G12 case, and the normalization factor varies by a factor of 2. Also in this case, the systematic uncertainty on the parameters due to the EBL models is larger than or equal to the statistical errors.

At $z=1.1$, D11 and F08 have similar attenuation values below \SI{200}{GeV}, which is the maximum energy in our analysis. The attenuation discrepancy between D11 and G12 is a factor of 2 at \SI{100}{GeV} and a factor of 5 at \SI{200}{GeV}. Thus, the intrinsic spectrum has a larger normalization and it is harder in the G12 case than in the D11 and F08 case, as seen in Table \ref{tab:Escale_EBL_systematics}.

The spectral index and normalization factor including the systematic uncertainties discussed above are therefore $-3.15^{+0.70}_{-0.70}\mathrm{(stat)}^{+0.98}_{-0.04}\mathrm{(sys)}^{+0.70}_{-0.04}\mathrm{(sysEBL)}$ and $2.03^{+0.39}_{-0.39}\mathrm{(stat)}^{+0.96}_{-0.89}\mathrm{(sys)}^{+1.96}_{-0.08}\mathrm{(sysEBL)}\times10^{-8}$\SI{}{\per\TeV\per\second\per\cm\squared} respectively. There are other systematic effects that may affect the spectral parameters (e.g. pointing accuracy and background uncertainty), as described in \cite{aleksic_2016}, however they are ignored in the analysis because they are less relevant than the aforementioned ones.

\subsection{Light Curve}
\label{subsec:energyLC}

\begin{figure}
\begin{center}
\includegraphics[width=0.9\columnwidth]{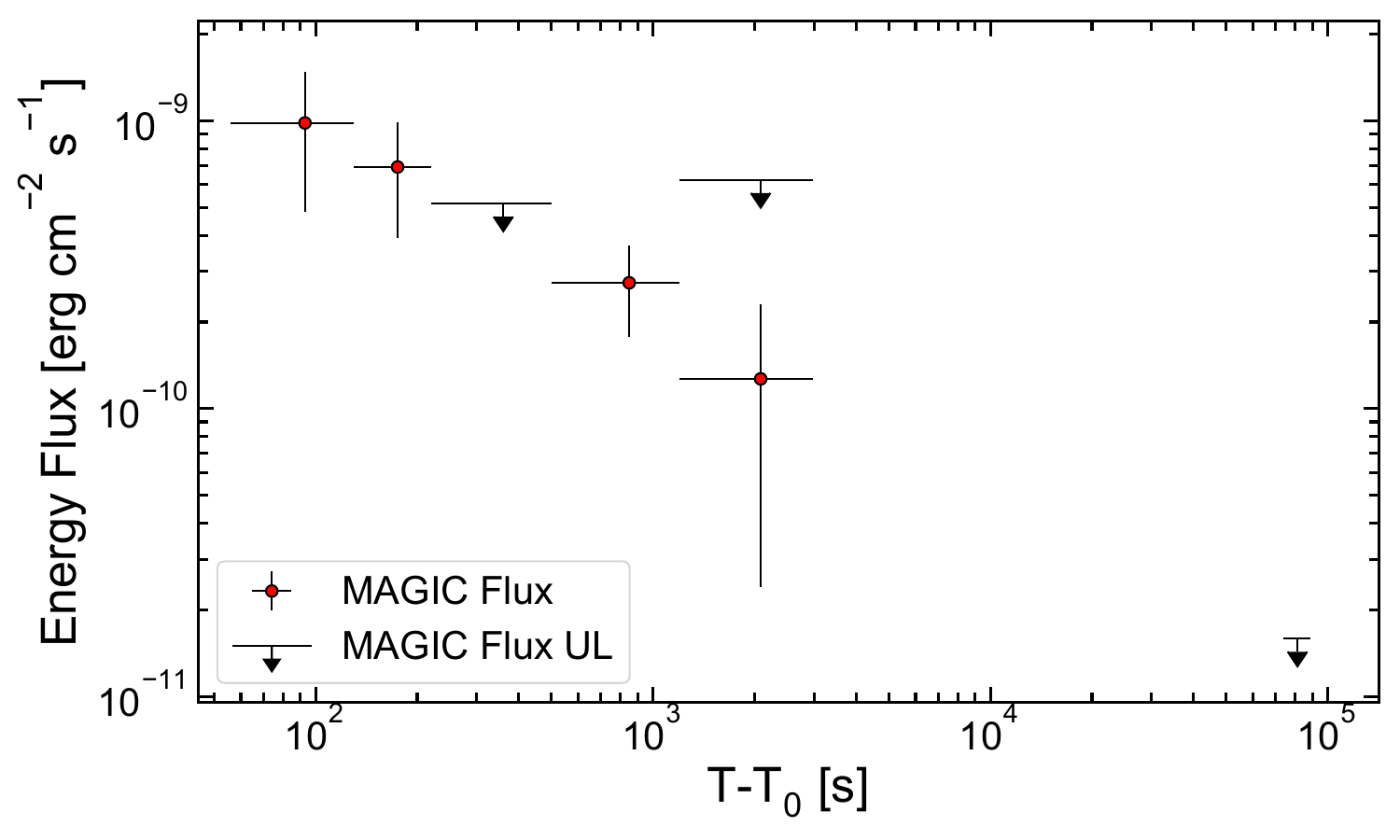}
\caption{EBL-corrected energy-flux light curve between \SI{70}{\GeV} and \SI{200}{\GeV} from $T_0+$\SI{56}{\second} to $T_0+$\SI{40}{\minute} (first night, divided into five time bins) and from $T_0+$\SI{20.5}{\hour} to $T_0+$\SI{24.6}{\hour} (second night). Upper limits are calculated as 95\% confidence level for the bins with relative errors >50\%.}
\label{fig:energy_light_curve}
\end{center}
\end{figure}

The VHE energy-flux light curve between \SI{70}{\GeV} and \SI{200}{\GeV} is shown in Fig.~\ref{fig:energy_light_curve}. The energy flux of each time bin is obtained by integrating the EBL-corrected forward-folded spectrum with the D11 model, so that the spectral variability with time is taken into account. For each time bin, the event cut is based on the signal survival fraction of simulated events in order to increase the statistics in such short time bins. The corresponding energy threshold is around 70 GeV for all the time bins. The light curve is compatible with a power-law decay. The best fit decay index until the 5th bin excluding upper limits is $-0.62\pm0.04$.

Upper limits are calculated for bins where relative flux errors are larger than 50\% using the method described in \citep{rolke}. The excess count upper limit of 95\% confidence interval is calculated for each of such bins and converted into the energy flux unit by assuming the power-law spectrum with an index of -3 attenuated with the D11 model.

The systematic uncertainties considered in Sec. \ref{subsec:spectrum} also affect the flux points in the light curve to a similar extent. However, since the spectral shape is not expected to change significantly during the short period of each bin of the light curve, the relative flux error is similar among all the bins. Therefore, the temporal decay index should be independent of the uncertainties as long as the spectrum is assumed to be stable. In fact, we could not detect any significant spectral changes larger than the statistical error during the time interval where the light-curve was produced.

From the analysis of the data on the second night, which spans from $T_0+$\SI{20.5}{\hour} to $T_0+$\SI{24.6}{\hour}, we found no significant excess around the position of the GRB with both the cut used in Sec. \ref{subsec:detection} and a conventional cut optimized for the Crab Nebula. 

We calculated the flux upper limit on the second night assuming an intrinsic power-law spectrum with an index of -3 and the EBL model D11. The event cut applied is the same one as used for the light curve on the first night. The upper limit of the EBL-corrected flux is shown in Fig.~\ref{fig:energy_light_curve}.
We note that the VHE luminosity of GRB~201216C implied by MAGIC observations is fainter (a factor ~10-30) than the luminosity predicted by \cite{zhang_ren23} on the basis of their afterglow modeling at lower frequencies.

\section{Multi-wavelength data from radio to gamma-ray}\label{sec:mwl}

In this Section we give an overview of the data at lower energies collected from the literature or analysed in this work, and later used for modeling and interpreting the overall emission (see Section~\ref{sec:modeling}). 
The data is shown are Figs.~\ref{fig:modeling_lc} and \ref{fig:modeling_sed}.

\subsection{Radio observations}
We collected radio observations from \cite{rhodes22}.
These late-time observations have been performed with e-MERLIN, the VLA, and MeerKAT, and cover the $\sim1-10$\,GHz frequency range. There are no simultaneous detections available at higher frequencies at the time of radio detections, which span the temporal window $\sim5-56$ days after $T_0$.
\cite{rhodes22} argue that the emission detected in the radio band is dominated by a different component as compared to the emission detected at earlier times in the optical band and in X-rays, and they suggest radiation from the cocoon as possible explanation.
In our analysis we also find that radio data cannot be easily explained as synchrotron radiation from the forward shock driven by the relativistic jet, see the discussion in Section~\ref{sec:modeling}. We nevertheless include radio data in our analysis (star symbols in Figs.~\ref{fig:modeling_lc} and \ref{fig:modeling_sed}), verifying that the estimation of the synchrotron flux from the jet given by the modeling lies below the observed radio emission.

\subsection{Optical observations: the Liverpool Telescope and VLT}

The 2-m fully robotic Liverpool Telescope (LT) autonomously reacted \citep{guidorzi2006} to the Swift-BAT alert, and started observations from about 178~s after the burst with the IO:O\footnote{\url{https://telescope.livjm.ac.uk/TelInst/Inst/IOO/}} optical camera in the SDSS-r band \citep{shrestha20}. The light curve of GRB~201216C optical counterpart initially displayed a  flat behaviour (see Fig.~\ref{fig:modeling_lc}) followed by a steepening, as revealed by VLT data gathered at 2.2 hours post-burst \citep{izzo2020} and by the non-detection of the afterglow in deeper LT observations at 1 day post-burst. We note that the LT photometry data was calibrated using a common set of stars present in the field of view selected from the APASS catalog.


\subsection{X-ray observations}
\textit{Swift}-XRT started to collect data on GRB~201216C only \SI{2966.8}{\second} after the burst onset due to an observing constraint (\citealt{grb201216c_swift_detection}). Observations continued up to $T_0+\SI{4325.4}{\second}$. 
The unabsorbed X-ray flux integrated in the 0.3-10\,keV energy range is shown in Fig.~\ref{fig:modeling_lc} (blue data points). \\
At around 0.1\,days XRT and optical data are simultaneously available and we built the spectral energy distribution (SED) around this time (Fig.~\ref{fig:modeling_sed}). The XRT spectrum has been derived by analysing data between 8900 and 9300\,s with the XSPEC software. Source and background spectra have been built using the automatic analysis tool\footnote{\url{https://www.swift.ac.uk/xrt\_spectra/01013243/}}. We modeled the spectrum with an absorbed power-law accounting both for Galactic and intrinsic metal absorption using the XSPEC models $tbabs$ and $ztbabs$, respectively.
The Galactic contribution is fixed to the value $N_{\rm H,G}=5.04\times10^{20}$\,cm$^{-2}$ \citep{willingale13}, while the column density in the host galaxy is a free parameter. 
We find that the best fit photon index is $-1.67\pm0.19$ and the intrinsic column density is $N_{\rm H}=(1.48\pm0.52)\times 10^{22}$\,cm$^{-2}$.
The spectral data, rebinned for plotting purposes and de-absorbed for both Galactic and intrinsic absorption, are shown in Fig.~\ref{fig:modeling_sed} (black crosses).

\subsection{Gamma-ray observations by Fermi-LAT}

\textit{Fermi}-LAT observations started from $T_0+\SI{3500}{\second}$ and continued until the GRB position was no longer visible ($T_0+\SI{5500}{\second}$).
No signal is detected during this time window.
Assuming a photon index $\alpha=-2$, the estimated upper limit in the energy range 0.1-1\,GeV is \SI{3e-10}{\erg\per\cm\squared\per\second}  \citep{grb201216c_fermilat}.
This upper limit is included in our analysis (orange arrow in Fig.~\ref{fig:modeling_lc}).
~\\

All the light curves at different frequencies are shown in Fig.~\ref{fig:modeling_lc}. 
The \textit{Swift}-BAT prompt emission light curve is also included in the figure (grey data points). 
The BAT flux is integrated in the 15-50\,keV energy range and points are rebinned using a signal-to-noise ratio (SNR) criterion equal to seven\footnote{\url{https://www.swift.ac.uk/burst\_analyser/01013243/}}.
The vertical colored stripes mark the times where SEDs are built. The SEDs are shown in Fig.~\ref{fig:modeling_sed}, where the MAGIC spectrum integrated between 56\,s and 1224\,s is also shown.


\section{Modeling}
\label{sec:modeling}
In this Section, we discuss the origin of the emission detected by MAGIC and its connection to the afterglow emission at lower energies, from radio to X-rays. 
In particular, we test an SSC scenario from electrons accelerated at the forward shock.
We consider a relativistic jet with initial Lorentz factor $\Gamma_0\gg1$, opening angle $\theta_{\rm jet}$, and a top-hat geometry. The (isotropic equivalent) kinetic energy of the jet $E_{\rm k}$ is related to $E_{\rm iso,\gamma}\sim6\times10^{53}$\,erg (see Sec.~\ref{sec:observation}) through the efficiency for production of prompt radiation $\eta_\gamma$: $E_{\rm k}=\frac{1-\eta_\gamma}{\eta_\gamma}E_{\rm iso,\gamma}$. 
The details of the equations adopted to describe the dynamics, the particle acceleration, and the radiative output can be found in \cite{miceli22} and are also reported in Appendix~\ref{app:numerical_model}. We summarize here the general model and the main assumptions.

The jet is expanding in an ambient medium characterised by a density described by a power-law function $n(R)\propto R^{-s}$. We consider the density to be either constant ($s=0$ and $n(R)=n_0$) or shaped by the progenitor stellar wind: $n(R)= A R^{-2}$ ($s=2$), where $A$ is related to the mass loss rate of the progenitor's star $\dot M$ and to the velocity of the wind $v_{\rm w}$ by $A =\dot M/4\,\pi\,m_{\rm p}\,v_w$ ($m_{\rm p}$ is the mass of the proton). We normalize the value of $A$ to a mass loss rate of $10^{-5}$ solar masses per year and a wind velocity of $10^3$\,km\,s$^{-1}$: $A=3\times10^{35}\,A_\star$\,cm$^{-1}$. We assume that ambient electrons are accelerated at the forward shock into a power-law distribution $dN/d\gamma\propto\gamma^{-p}$ from $\gamma_{\rm min}$ and $\gamma_{\rm max}$. 
The bulk Lorentz factor of the fluid just behind the shock is assumed to be constant ($\Gamma=\Gamma_0$) before the deceleration and described by the solution given by \cite{bm76} ($\Gamma=\Gamma_{\rm BM}$) during the deceleration (note that the equation given by \cite{bm76} describes the Lorentz factor of the shock $\Gamma_{\rm sh}$, which we relate to the Lorentz factor of the fluid using $\Gamma=\Gamma_{\rm sh}/\sqrt{2}$). The two regimes are smoothly connected to obtain the description of the bulk Lorentz factor of the fluid just behind the shock as a function of shock radius.

To infer the particle distribution and the photon spectrum at any time $t$ we numerically evolve the equations describing the electron and photon populations including adiabatic losses, synchrotron emission and self-absorption, Inverse Compton emission and $\gamma-\gamma$ annihilation and pair production. To relate the comoving properties computed by the code to the observed one, we assume that the emission received at a given observer time is dominated by electrons moving at an angle $\cos\theta=\beta$ from the line of sight to the observer, where $\beta$ is the velocity of the shocked fluid. 

\begin{figure}
\begin{center}
\includegraphics[width=1\columnwidth]{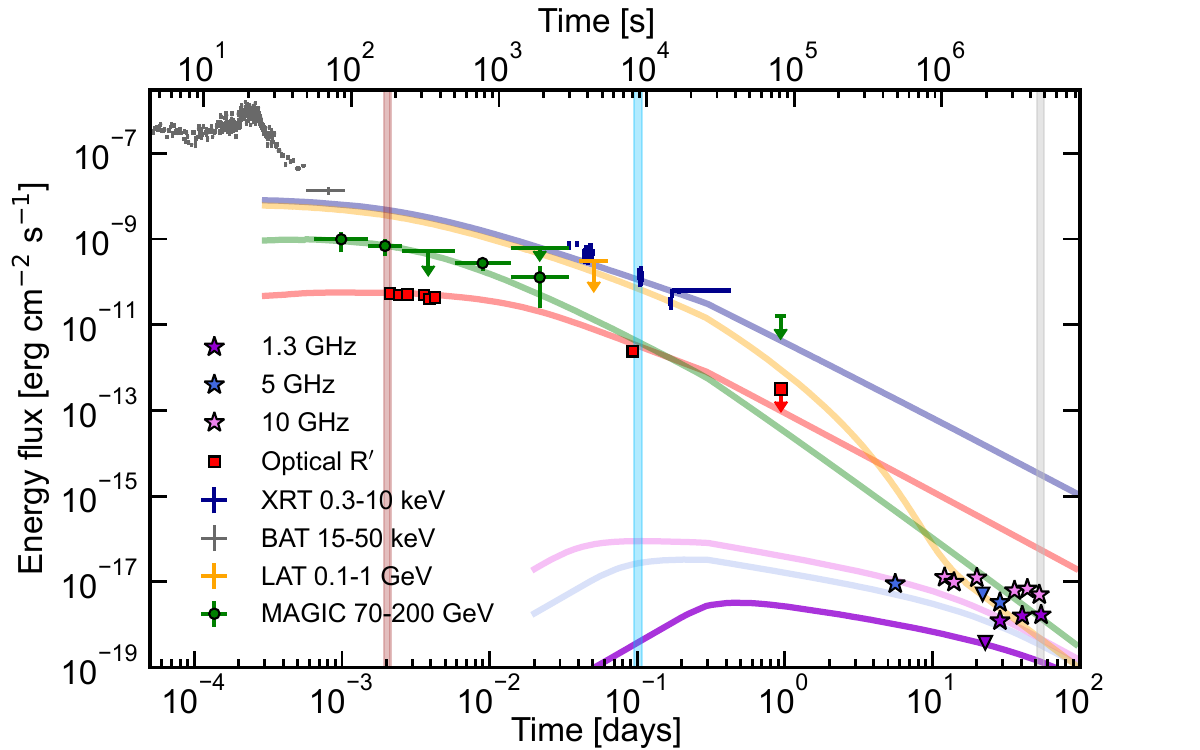}
\caption{Multi-wavelength light curves of GRB~201216C. Both the X-ray and optical observations have been corrected accounting for absorption. MAGIC data points are EBL-corrected. Upside down triangles represent upper limits. Solid curves show the best fit model obtained in a synchrotron - SSC forward shock scenario. Different colors refer to the different wavelengths where observations are available (see the legend). The modeling is obtained with the following parameters: $E_{\rm k}=4\times10^{53}$\,erg, $\epsilon_{\rm e}=0.08$, $\epsilon_{\rm B}=2.5\times10^{-3}$, $A_\star=2.5\times10^{-2}$, $p=2.1$, $\Gamma_0=180$, and $\theta_{\rm jet}=1^\circ$. Vertical lines mark the times where SED have been built (see Fig.~\ref{fig:modeling_sed}).}
\label{fig:modeling_lc}
\end{center}
\end{figure}

\begin{figure}
\begin{center}
\includegraphics[width=1\columnwidth]{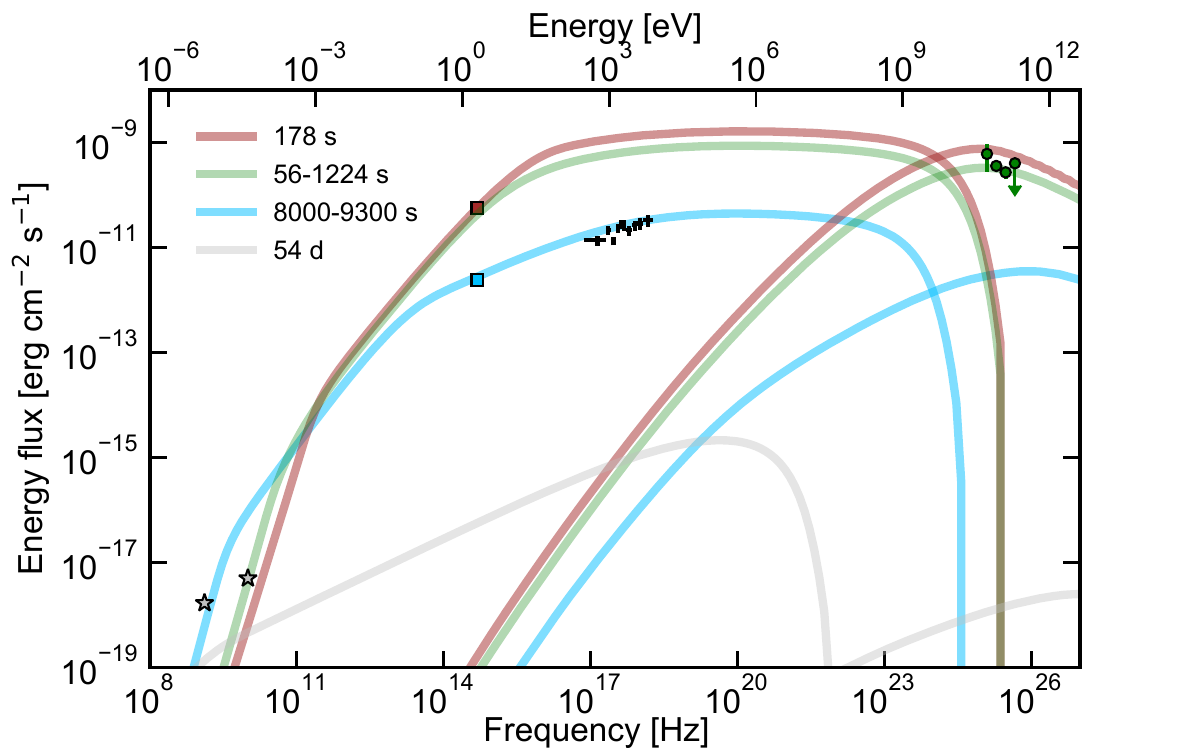}
\caption{SEDs of GRB~201216C at different times. Different colors for curves and data points refer to different times (see the legend). The times where the SEDs are calculated are also marked in Fig.~\ref{fig:modeling_lc} with vertical stripes. Solid curves show the synchrotron and SSC theoretical spectra for the same parameters used for Fig.~\ref{fig:modeling_lc}. 
De-absorbed optical data in the $r'$ filter are marked with square symbols, while star symbols are observations at 1.3\,GHz and 10\,GHz at 54.5\,d and 53\,d, respectively (from \citealt{rhodes22}). The XRT spectral data points estimated around 9000\,s are also shown.
Green circles show the MAGIC spectrum averaged between 56 and 1224\,s (Fig.~\ref{fig:time_integrated_spectrum}). 
The theoretical SED to be compared with the MAGIC spectrum is the green curve, which shows the predicted spectrum (synchrotron + SSC) averaged in the same time window (56-1224\,s).}
\label{fig:modeling_sed}
\end{center}
\end{figure}

Before presenting the results of the numerical modeling, we discuss some general considerations that can be inferred using analytic approximations from \cite{gs02}.
Fig.~\ref{fig:modeling_lc} shows that the optical flux is nearly constant up to at least $5\times10^{-3}$\,d. At later times this behaviour breaks into a steeper temporal decay. We take as reference value for the break time $\sim10^{-2}$\,d. This behavior of the optical light curve can be explained if the break frequency $\nu_{\rm m}$ (i.e. the typical photon energy emitted by electrons with Lorentz factor $\gamma_{\rm min}$) is crossing the $r$ band. The nearly constant flux before the crossing time is indicative of a wind-shaped external  medium (i.e., $s=2$). The preference for a wind-like medium is also supported by the lack of a phase of increasing flux in the MAGIC observations, which start as early as $\sim$60\,s after the onset of the prompt emission. Since $\nu_{\rm m}\propto t^{-1.5}$, we expect $\nu_{\rm m}\sim1$\,GHz at $\sim50$\,d.
Observations at 1.3\,GHz do not allow to constrain the peak time, but we notice that they are consistent with the presence of a peak around 50\,days (a zoom on radio observations can be found in figure 1 of \citealt{rhodes22} and it shows that, considering the errors, the flux is consistent with being constant at about 50\,days). The radio SED at this time (see Fig.~\ref{fig:modeling_sed} and also \citealt{rhodes22}) shows that the self-absorption frequency $\nu_{\rm sa}$ must be below 1\,GHz. Since $\nu_{\rm sa}\propto t^{-3/5}$, this implies that during the time spanned by observations $\nu_{\rm sa}<\nu_{\rm m}$. The fast increase of the 1.3\,GHz flux (with temporal index $\gtrsim 5$, as reported in \citealt{rhodes22}) however implies that observations at this time are below the self-absorption frequency (otherwise the flux at 1.3\,GHz should be constant), constraining $\nu_{\rm sa}(50\,\rm{d})\sim$1\,GHz.

To summarise, the scenario implied by optical and radio observations invokes a jet expanding in a wind-like density and producing a synchrotron spectrum with $\nu_{\rm sa}<\nu_{\rm m}$, and $\nu_{\rm m}$ crossing the optical $r$ band at $\sim$\,$10^{-2}$\,d and the 1\,GHz frequency at $\sim50$\,d. We now check the consistency of this interpretation with X-ray observations. Imposing $\nu_{\rm m}(54\,\rm{d})=\nu_{\rm sa}(54\,\rm{d})=1$\,GHz and the flux $F(\nu_{\rm sa}, 54\,{\rm d})=2\times10^{-18}$\,erg\,cm$^{-2}$\,s$^{-1}$, and using equations for the break frequencies and flux in a wind-like medium from \cite{gs02}, it is possible to derive the values of $E_{\rm k}, \epsilon_{\rm B}$, and $A_\star$ as a function of $\epsilon_{\rm e}$, for fixed values of $p$. For $p=2.2$ we find $E_{\rm k,52}\simeq\epsilon_{\rm e}$, $\epsilon_{\rm B}\simeq2.3\times10^{-6}\epsilon_{\rm e}^{-5}$ and $A_\star\simeq8.8\,\epsilon_{\rm e}^{2}$. This shows that the requirement that the $F_{\nu}$ spectrum peaks at $\nu_{\rm sa}=1$\,GHz at 54 days limits the energy to a low value $E_{\rm k}<10^{52}$\,erg, inconsistent with the large flux detected in the X-ray afterglow. In particular we find that the X-ray band is always above the cooling frequency $\nu_{\rm c}$ for different assumptions on $\epsilon_{\rm e}$. Moreover, in this range, the predicted flux is at least one order of magnitude below the detected X-ray flux. This statement is quite robust, as the flux in this band weakly depends on $\epsilon_{\rm B}$, does not depend on $A_\star$, and is proportional to $E_{\rm k}\,\epsilon_{\rm e}$. Pushing $\epsilon_{\rm e}$ to large values (close to one) improves the situation, at the expense of a very small $\epsilon_{\rm B}$, implying a large SSC component. This solution is ruled out by MAGIC observations.

Being unable to find a scenario that explains all the available data as synchrotron and SSC emission from the forward shock driven by a relativistic jet, we consider the possibility that late-time radio emission is dominated by a different component, as also concluded by \cite{rhodes22}, which identify in a wider mildly (or non) relativistic cocoon the origin of the radio emission. We then restrict the modeling to the MAGIC, X-ray and optical data, requiring that the flux  at 1-10\,GHz from the narrow relativistic jet is below the observed flux.


We performed numerical calculations of the expected synchrotron and SSC radiation and their evolution in time for wide ranges of values of the parameters $E_{\rm k}, \epsilon_{\rm e}, \epsilon_{\rm B}, p, n(R), \theta_{\rm jet}, \rm{and}~\Gamma_0$. The investigated range of values for each parameter is reported in Table~\ref{tab:model_parameters}. 
The numerical calculations confirm the considerations derived from analytic estimates. In particular, we neither find a solution for a homogeneous medium nor for a complete description of radio to GeV observations.
Assuming a wind-like density profile, we find that the observations can be well described as synchrotron and SSC radiation. 
In particular, once the request to model also radio observations with forward shock emission from the relativistic jet is abandoned, the X-ray flux can be explained by increasing the assumed value of the jet energy, which also moves the self-absorption frequency to lower energies. 
An example of modeling is provided in Figs.~\ref{fig:modeling_lc} and \ref{fig:modeling_sed}, where observations (corrected for absorption in the optical and X-ray band) are compared to the light curves and spectra predicted with the following parameters: $E_{\rm k}=4\times10^{53}$\,erg, $\epsilon_{\rm e}=0.08$, $\epsilon_{\rm B}=2.5\times10^{-3}$, $A_\star=2.5\times10^{-2}$, $p=2.1$, $\Gamma_0=180$, and $\theta_{\rm jet}=1^\circ$ the values are listed also in Table~\ref{tab:model_parameters}.
The jet opening angle is broadly constrained by the need to not overproduce the radio flux. The inferred value points to a narrow jet, with opening angle in the low-value tail of distributions of inferred jet opening angles for long GRBs \citep{chen20}.
We note that a similarly small ($\theta_{\rm jet}\sim$\,0.8$^\circ$) value for the jet opening angle has been inferred for the TeV GRB~221009A \cite{lhaaso}.
The inferred jet kinetic energy implies an efficiency of the prompt emission $\eta_\gamma\simeq60\%$. In agreement with the steep optical spectrum reported by \cite{grb201216c_redshift}, this model implies an extinction of 4.6 magnitudes in the $r'$ band, which is well in excess of the Galactic contribution ($E(B-V)=0.05$). As it can be seen from Fig.~\ref{fig:modeling_lc}, the onset of the deceleration occurs at $t_{\rm obs}\lesssim200$\,s, where the X-ray and TeV theoretical light curves steepen from an almost flat to a decaying flux. 

The steepening of the optical light curve instead occurs at $\sim10^3$\,s because, as already commented, it is determined by the $\nu_{\rm m}$ frequency crossing the $r$ band.
In this interpretation, the frequency $\nu_{\rm m}$ is initially above the optical band (see the brown SED in Fig.~\ref{fig:modeling_sed}) and then moves to lower frequencies crossing the optical and explaining the steepening in the light curve. X-ray observations lie just above the cooling frequency, but the X-ray spectrum remains harder than expected due to the role of the Klein-Nishina cross section.
We also computed the expected SED averaged between 56\,s and 1224\,s, where the MAGIC spectrum (see Fig.~\ref{fig:time_integrated_spectrum}) is computed. The model SED is reported in Fig.~\ref{fig:modeling_sed} (green curve, to be compared with the MAGIC data, green circles). We find that the $\gamma-\gamma$ internal absorption plays a minor role in shaping the spectrum: the flux reduction at 200\,GeV is about 25\%.
In the same figure it is also possible to see the expected location of the maximum energy of synchrotron photons, 
initially located at 10\,GeV at the time of the first SED, and then moving towards lower energies. Assuming diffusive shock acceleration proceeding at the maximum rate rules out a synchrotron origin for the photons detected by MAGIC.

\begin{table}
\begin{center}
\begin{tabular}{c|c|c}
    Parameter & Range & Best fit value\\
    \hline\hline
    $E_{\rm k}$ [erg] & $10^{50}-10^{54}$ & $4\times10^{53}$ \\
    \hline
    $\theta_{\rm jet}$ [degrees] & $0.5-3$ & 1 \\
    \hline
    $\Gamma_{\rm 0}$ & 80-300 & 180 \\
    \hline    $n_0$ [cm$^{-3}$] ($s=0$) & $10^{-2}-10^{2}$ &  - \\
    \hline
    $A_{\star}$ ($s=2$) & $10^{-2}-10^{2}$ & $2.5\times10^{-2}$ \\
    \hline
    $p$ & 2.05 - 2.6 & 2.1 \\
    \hline
    $\epsilon_{\rm e}$ & 0.01-0.9 &  0.08\\
    \hline
    $\epsilon_{\rm B}$ & $10^{-7}-10^{-1}$ & $2.5\times10^{-3}$ \\
\end{tabular}
\caption{List of the input parameters for the afterglow model. For each parameter, the range of values investigated by means of the numerical model are listed in the second column. Solutions are not found for an homogeneous density medium ($s=0$). The last column list the values that better fit the observations and used to produce the model light-curves and model SEDs in Figs.~\ref{fig:modeling_lc} and \ref{fig:modeling_sed}.}
\label{tab:model_parameters}
\end{center}
\end{table}


\section{Conclusions}
\label{sec:conclusions}

In this paper, MAGIC analysis results on GRB~201216C and their interpretation were presented. The GRB afterglow was observed at early times ($\sim 10^2-10^3$\,s) by MAGIC for a total of $\sim$\SI{2.5}{\hour} during the first night and detected at the level of $6\sigma$ in the first 20 minutes. This is the second firm detection of a GRB with the MAGIC telescopes after GRB~190114C, and also the farthest VHE source detected to date. Both the observed and intrinsic average spectra can be well described by a power-law. A time-resolved analysis was also performed, in order to evaluate the temporal behavior of VHE emission. The obtained light curve shows a monotonic power-law decay, indicating a probable afterglow origin of the VHE emission.

Multi-wavelength data were also collected by other ground and space-based instruments. Unfortunately, most of them are not contemporaneous to the first MAGIC observation time window. In other cases, as for \textit{Fermi}-LAT, the GRB could not be detected. Like other GRBs detected in the VHE range, multi-wavelength data was used to perform a modeling of the broadband emission. In this manuscript a synchrotron and SSC radiation model at the forward shock in the afterglow was considered. SEDs built at different times show that synchrotron photons can reach a maximum energy of \SI{10}{\GeV} about three minutes after the GRB onset. The emission detected by MAGIC reaches higher energies, and can therefore be explained by the SSC component of the model. By comparing analytic estimates and the numerical modeling, evidence for the need of a different component at the origin of late-time radio emission is found, in agreement with previously published studies on this GRB. Both observations and modeling support a wind-like medium, as expected in the case of a long GRB. The best fit model parameters are found to be consistent with those estimated in previous studies of GRB afterglows without VHE detection. This proves the flexibility of the SSC scenario in describing the VHE emission of GRBs.

Like other VHE detected GRBs (GRB~180720B and GRB~190114C), 201216C was a bright GRB, allowing for a detection in spite of the high redshift. Once again, the rapid response and low-energy threshold of the MAGIC telescopes to GRB alerts was crucial to detect the VHE emission in the early afterglow phase. Altogether, the detection by MAGIC and other experiments of several bursts so far suggests that VHE emission is common both in high and low-luminosity GRBs. Other VHE detected GRBs showed a correlation between the intrinsic emission in the X-ray and VHE bands, where a similar time decay and flux value were observed. In the case of GRB~201216C such a direct comparison cannot be performed given the lack of contemporaneous data in the two bands. The extrapolation of the X-ray flux into the first MAGIC time window, assuming a smooth power-law behavior typical of the afterglow phase, shows that the VHE flux is lower than the X-ray one. However, one should take into account the rather narrow energy range of the VHE detection due to the large absorption caused by the EBL.

\section*{Acknowledgements}

The MAGIC collaboration would like to thank the Instituto de Astrof\'{\i}sica de Canarias for the excellent working conditions at the Observatorio del Roque de los Muchachos in La Palma. The financial support of the German BMBF, MPG and HGF; the Italian INFN and INAF; the Swiss National Fund SNF; the grants PID2019-104114RB-C31, PID2019-104114RB-C32, PID2019-104114RB-C33, PID2019-105510GB-C31, PID2019-107847RB-C41, PID2019-107847RB-C42, PID2019-107847RB-C44, PID2019-107988GB-C22, PID2020-118491GB-I00 funded by the Spanish MCIN/AEI/ 10.13039/501100011033; the Indian Department of Atomic Energy; the Japanese ICRR, the University of Tokyo, JSPS, and MEXT; the Bulgarian Ministry of Education and Science, National RI Roadmap Project DO1-400/18.12.2020 and the Academy of Finland grant nr. 320045 is gratefully acknowledged. This work was also been supported by Centros de Excelencia ``Severo Ochoa'' y Unidades ``Mar\'{\i}a de Maeztu'' program of the Spanish MCIN/AEI/ 10.13039/501100011033 (SEV-2016-0588, CEX2019-000920-S, CEX2019-000918-M, CEX2021-001131-S, MDM-2015-0509-18-2) and by the CERCA institution of the Generalitat de Catalunya; by the Croatian Science Foundation (HrZZ) Project IP-2016-06-9782 and the University of Rijeka Project uniri-prirod-18-48; by the Deutsche Forschungsgemeinschaft (SFB1491 and SFB876); the Polish Ministry Of Education and Science grant No. 2021/WK/08; and by the Brazilian MCTIC, CNPq and FAPERJ.
This work was supported by JSPS KAKENHI Grant Number JP21K20368. L. Nava acknowledges partial support from the INAF Mini-grant 'Shock acceleration in Gamma Ray Bursts'. \\
The Liverpool Telescope is operated on the island of La Palma by Liverpool John Moores University in the Spanish Observatorio del Roque de los Muchachos of the Instituto de Astrofisica de Canarias with financial support from the UK Science and Technology Facilities Council (STFC) under UKRI grant ST/T00147X/1. Manisha Shrestha and Iain Steele thank UKRI/STFC for financial support (ST/R000484/1). \\
We would like to thank the STARGATE collaboration for the confirmation of the redshift of the source via private communication.

\section*{Author Contributions}
K.~Asano gave support for the interpretation and modeling of the MAGIC and multi-wavelength data. \\
A.~Berti coordinated the paper project and contributed to the paper drafting. \\
S.~Fukami contributed to the MAGIC data analysis and the paper drafting. \\
S.~Loporchio provided MAGIC data analysis, paper drafting, and discussion. \\
L.~Nava provided the main interpretation and modeling of the MAGIC and multi-wavelength data, and contributed to the paper drafting.  \\
Y.~Suda provided MAGIC data analysis and paper drafting. \\
A.~Melandri contributed to the Liverpool Telescope IO:O Camera data analysis.\\
C. G.~Mundell: leadership of the Liverpool GRB group, followup development and contributions to GRB science.
M.~Shrestha contributed to the Liverpool Telescope IO:O Camera data analysis.\\
I. A.~Steele led the development and initial calibration of the Liverpool Telescope IO:O Camera and the GRB followup software.  \\
The rest of the authors have contributed in one or several of the following ways: design, construction, maintenance and operation of the instrument(s)
used to acquire the data; preparation and/or evaluation of the observation proposals; data acquisition, processing, calibration and/or reduction; production of analysis tools and/or related Monte Carlo simulations; overall discussions about the contents of the draft, as well as related refinements in the descriptions.

\section*{Data Availability}

The data underlying this article will be shared on reasonable request to the corresponding authors.



\bibliographystyle{mnras}
\bibliography{grb201216c_biblio} 




\appendix
\label{appendix}
\section{Numerical afterglow model}
\label{app:numerical_model}
In this section, we summarise the equations adopted to describe the synchrotron and synchrotron-self Compton (SSC) emission from electrons accelerated in the forward shock driven by a relativistic jet ($\Gamma\gg1$) in the surrounding medium.
The list of the free model parameters can be found in Table~\ref{tab:model_parameters}. More details can be found in \cite{miceli22}.

\subsubsection*{Circumburst medium}
The surrounding medium is assumed to have a radial density profile $n(R)$ described by a power-law function: $n(R)=A R^{-s}$, where $R$ is the distance of the shock front from the center of the explosion and $n$ is the number density.
Two possibilities are investigated: $s=0$ ($n(R)=\textup{constant}$) and $s=2$ ($n(R)=A\,R^{-2}$). In the latter case, the normalization $A$ is related to the mass loss rate of the progenitor’s star $\dot M$ and to the velocity of the wind $v_{\rm w}$
by $A=\dot M/4\,\pi\,m_{\rm p}\,v_{\rm w}$ ($m_{\rm p}$ is the proton mass). We normalize
the value of $A$ to a mass loss rate of $10^{-5}$ solar masses per year and a wind velocity of $10^3$\,km\,s$^{-1}$: $A = 3\times10^{35}A_\star$\,cm$^{-1}$.

\subsubsection*{Blast-wave dynamics}
The evolution of the Lorentz factor of a decelerating relativistic adiabatic blastwave has been derived by \cite{bm76} (BM), which provides the shock Lorentz factor, its relation with the Lorentz factor of the shocked fluid just behind the shock, and the profile of the Lorentz factor in the downstream region. We consider the BM solution to describe the Lorentz factor of the shock $\Gamma_{\rm sh}$:
\begin{equation}
    \Gamma_{\rm sh,BM} (R) = \left[\frac{(17-4s)\,E_{\rm k}}{8\pi\,A\,m_{\rm p}c^2\,R^{(3-s)}}\right]^{1/2}~,
\end{equation}
The Lorentz factor of the fluid just behind the shock ($\Gamma$) is given by:
\begin{equation}
    \Gamma = \frac{\Gamma_{\rm sh}}{\sqrt{2}}~.
\end{equation}
We do not take into account its profile in the downstream region (homogeneous shell approximation).

The BM solution is valid only during the deceleration, in particular when the energy of the ejecta is negligible compared to the energy transferred to the shocked external medium. Well before the deceleration begins, the Lorentz factor is constant and equal to its initial value $\Gamma_0$.
To describe the entire evolution, we smoothly connect the initial phase to the deceleration phase:
\begin{equation}
    \Gamma = \Gamma_0\left[1+ \left(\frac{\Gamma_0}{\Gamma_{\rm BM}}\right)^q\right]^{-1/q}~.
\end{equation}
For fitting purposes, in GRB~201216C the smoothing factor $q$ has been kept fixed to the value $q=3$ to describe the transition from the coasting to the deceleration phase.

\subsubsection*{Particle acceleration and magnetic field acceleration}
The accelerated electrons are assumed to have a spectrum described by a power-law: $dN^{\rm acc}(\gamma)/ d\gamma \propto \gamma^{-p}$ for $ \gamma_{\rm min} \leq \gamma \leq \gamma_{\rm max}$ where $\gamma_{\rm min}$ and $\gamma_{\rm max}$ are the minimum and maximum Lorentz factors at which electrons are accelerated. 

The value of  $\gamma_{\rm max}$ is obtained by imposing that the synchrotron cooling time is equal to the acceleration time, set by the assumption that the particles mean free path is equal to their Larmor radius. This leads to:
\begin{equation}
    \gamma_{\rm max} = \sqrt{\frac{6\pi\,q}{\sigma_{\rm T}B}}~.
    \label{eq:gamma_max}
\end{equation}
For a discussion on the maximum energy, see also \citep{kumar_gmax,derishev}, which use the same or  similar equations, differing at most by a numerical factor of the order of unity.

Since a fraction $\epsilon_e$ of the shock-dissipated energy goes into the acceleration of electrons into a non-thermal distribution, their average random Lorentz factor $\langle\gamma\rangle$ is: 
\begin{equation}
    \langle \gamma\rangle = \epsilon_{\rm e} \frac{m_{\rm p}}{m_{\rm e}} (\Gamma - 1)~,
    \label{eq:gamma_medium_shock}
\end{equation}
the minimum Lorentz factor is found after numerically solving the equation:
\begin{equation}
    \Bigg[\frac{\gamma_{\rm min}^{-p+2} - \gamma_{\rm max}^{-p + 2}}{\gamma_{\rm min}^{-p+1} - \gamma_{\rm max}^{-p + 1}} \Bigg] = \epsilon_e \frac{m_{\rm p}}{m_{\rm e}}\frac{p - 2}{p - 1} (\Gamma - 1) \hspace{1.cm} \text{if $p \neq 2$~.}
    \label{eq:gamma_min_implicit}
\end{equation}
For $\gamma_{\rm max}^{-p+2} \ll \gamma_{\rm min}^{-p+2}$, the equation reduces to the widely used one:

\begin{equation}
    \gamma_{\rm min} = \epsilon_{\rm e} \frac{m_{\rm p}}{m_{\rm e}}\frac{p-2}{p-1}(\Gamma-1)~.
\end{equation}
Because we find that the best fit of GRB~201216C requires $p\sim2.1$, we use the full equation for the estimate of $\gamma_{\rm min}$. Using the approximated one would result in underestimating $\gamma_{\rm min}$ and in the non-conservation of the number and total energy of the accelerated electrons.

A fraction $\epsilon_B$ of the shock-dissipated energy is used to amplify the magnetic field, giving:
\begin{equation}
    B = \sqrt{32\pi\,\epsilon_B m_{\rm p}c^2\,n(r)}\,\Gamma
\end{equation}

\subsubsection*{The evolution of the electron distribution}

The temporal evolution of the particle distribution $N(\gamma,t')$ as a function of the electron Lorentz factor $\gamma$ and the comoving time $t'$ is described by the differential equation:
\begin{equation}
    \frac{\partial N(\gamma,t')}{\partial t'} = \frac{\partial}{\partial\gamma}\bigg[\dot{\gamma}N(\gamma,t')\bigg] + Q(\gamma)~,
    \label{eq:continuity_equation}
\end{equation}
where $\dot{\gamma} = \partial\gamma/\partial t'$ is the rate of change of the Lorentz factor $\gamma$ of an electron caused by adiabatic, synchrotron and SSC losses and by energy gains due to absorption of synchrotron photons (synchrotron self-absorption, SSA).  

The source term $Q(\gamma,t')=Q^{\rm acc}(\gamma,t')+Q^{\rm pp}(\gamma,t')$ describes the injection of freshly accelerated particles ($Q^{\rm acc}(\gamma,t')=dN^{\rm acc}/d\gamma\,dt'$) and the injection of pairs $Q^{\rm pp}(\gamma,t')$ produced by photon-photon annihilation.

To solve the equation, we adopt an implicit finite difference scheme based on the discretization method proposed by \cite{chang_cooper}.

\subsubsection*{Electron energy losses}
The synchrotron power emitted by an electron with Lorentz factor $\gamma$ depends on the pitch angle, i.e., the angle between the electron velocity and the magnetic field line.
We assume that the electrons have an isotropic pitch angle distribution and use equations that are averaged over the pitch angle. The synchrotron cooling rate of an electron with Lorentz factor $\gamma$ is given by:
\begin{equation}
    \dot\gamma_{\rm syn} \equiv \frac{d\gamma}{dt'}\bigg|_{\rm syn} = -\frac{\sigma_T \gamma^2 B'^2}{6\,\pi\,m_{\rm e} c}~.
\end{equation}

The energy loss term for the SSC is calculated with the equation:
\begin{equation}
    \dot{\gamma}_{\rm SSC} = \frac{d\gamma}{d t'}\bigg|_{\rm SSC} = - \frac{3 h \sigma_t}{4 m_{\rm e} c \gamma^2} \int d\nu' \nu' \int \frac{d\Tilde{\nu'}}{\Tilde{\nu'}} n_{\Tilde{\nu'}} (t') K(\gamma,\nu',\Tilde{\nu'})~,
    \label{eq:energy_losses_IC}
\end{equation}
where $\Tilde{\nu'}$ and $\nu'$ are the frequencies (in the comoving frame) of the photon before and after the scattering, respectively.
For the expression of $K(\gamma,\nu',\Tilde{\nu'})$ we adopt the formulation proposed in \cite{jones}, which is valid both in Thomson and Klein-Nishina regime, and describes both the down-scattering (i.e. $\nu' < \Tilde\nu'$) and the up-scattering (i.e. $\nu' > \Tilde\nu'$) process.

Particles loose their energy also adiabatically, due to the spreading of the emission region:
\begin{equation}
    \dot{\gamma}_{\rm ad} = \frac{d\gamma}{d t'}\bigg|_{\rm ad}=-\frac{\gamma\beta^2}{3}\frac{d\ln V'}{dt'}~.
\end{equation}
The comoving volume $V'$ of the emission region can be estimated considering that the contact discontinuity is moving away from the shock at a velocity $c/3$ \citep{pennanen}. After a time $t'=\int dR/\Gamma(R)\,c$ the comoving volume is: 
\begin{equation}
    V' = 4 \pi R^2 \frac{ct'}{3}~,
\end{equation}

\subsubsection*{Estimate of the radiative output}
Following \cite{ghisellini_svensson}, the synchrotron spectrum emitted by an electron with Lorentz factor $\gamma$, averaged over an isotropic pitch angle distribution is:
\begin{equation}
\begin{split}
P'^{\rm syn}_{\nu'} (\nu',\gamma) = &\frac{2\,\sqrt{3}\,e^3\,B'}{m_{\rm e}\,c^2}\,x^2~\times\\ 
\times& \left[K_{4/3}(x)\,K_{1/3}(x) - 0.6x(K^2_{4/3}(x)-K^2_{1/3}(x))\right] ~,
    \label{eq:power_syn}
\end{split}
\end{equation}
where $x\equiv\nu'\,4\pi\,m_{\rm e}c/(6\,q\,B'\gamma^2)$, and $K_{\rm n}$ are the modified Bessel functions of order $n$.
The total power emitted at the frequency $\nu'$ is obtained integrating over the electron distribution:
\begin{equation}
    P'^{\rm syn}_{\nu'} (\nu') = \int{P'^{\rm syn}_{\nu'} (\nu',\gamma) \frac{dN}{d\gamma}d\gamma}  ~. 
\end{equation}

The SSC radiation emitted by an electron with Lorentz factor $\gamma$ can be calculated as:
\begin{equation}
    P'^{\rm SSC}_{\nu'} (\nu',\gamma) = \frac{3}{4} h \sigma_{\rm T} c \frac{\nu'}{\gamma^2} \int \frac{d\Tilde{\nu'}}{\Tilde{\nu'}} n_{\Tilde{\nu'}} K(\gamma,\nu',\Tilde{\nu'})~,
    \label{eq:power_IC}
\end{equation}
where $n_{\Tilde{\nu'}}$ is the photon density of synchrotron photons and the integration is performed over the entire synchrotron spectrum. Integration over the electron distribution provides the total SSC emitted power at frequency $\nu'$.

\subsubsection*{Absorption processes}
Electrons can re-absorb low energy photons before they escape from the source region. The cross section of the process is \citep{rybicki_lightman}:
\begin{equation}
    \sigma(\nu',\gamma) = - \frac{1}{8\pi\nu'^2m_{\rm e}}  \frac{P'(\gamma,\nu')\gamma^2}{N(\gamma)}\frac{\partial}{\partial\gamma}\bigg[\frac{N(\gamma)}{\gamma^2}\bigg]
    \label{eq:SSA_absorption}
\end{equation}
valid for any radiation mechanism at the emission frequency $\nu'$, with $P'(\gamma,\nu')$ being the specific power of electrons with Lorentz factor $\gamma$ at frequency $\nu'$ and assuming $h\nu' \ll \gamma m_{\rm e} c^2$. 

While the SSA mechanism will affect mostly the low frequency range, at the highest energies the flux can be attenuated by photon-photon annihilation.
For the cross section $\sigma(\nu'\nu'_t)$ (where $n'(\nu'_t)$ is the number density of the target photons) we use the equation:
\begin{equation}
    \sigma(\nu',\nu'_t) = \frac{3}{16}\sigma_T (1 - \beta'^2) \Bigg[(3 - \beta'^4)\ln{\bigg(\frac{1 + \beta'}{1 -\beta'}\bigg) - 2\beta'(2 - \beta'^2 ) \Bigg]}
\end{equation}
where:
\begin{equation}
    \beta'(\omega_t,\omega_s,\mu) = \Bigg[ 1 - \frac{2}{\omega_t \omega_s (1 - \mu)}\Bigg]^{\frac{1}{2}}
\end{equation}
and $\omega_t = h \nu'_t /m_e c^2$ with $\nu'_t$ being the target photon frequency, $\omega_s = h \nu' /m_e c^2$ with $\nu'$ being the source photon frequency and $\mu = \cos{\phi}$, where $\phi$ is the scattering angle. Then, it is possible to derive the annihilation rate of photons into electron-positron pairs as:
\begin{equation}
    R (\omega_t,\omega_s) = c \int_{-1}^{\mu_{max}} \frac{d\mu}{2} (1 - \mu) \sigma_{\gamma \gamma} (\omega_t,\omega_s,\mu)~,
    \label{eq:exact_pair_annihilation}
\end{equation}
where $\mu_{max} = \max(-1, 1-2/\omega_s \omega_t)$ coming from the requirement $\beta'^2 > 0$. Considering $x = \omega_t \omega_s$ it is possible to derive asymptotic limits for $R(\omega_t, \omega_s) \equiv R(x)$ in two regimes. For $x \to 1$ (i.e. near the threshold condition) $R(x) \to c \sigma_T/2 (x-1)^{3/2}$, while for $x \gg 1$ (i.e. ultra-relativistic limit) $R \to \frac{3}{4}c \sigma_T \ln{x}/x $. An accurate and simple approximation which takes into account both regimes is given by:
\begin{equation}
    R(x) \approx 0.652 c \sigma_T \frac{x^2 - 1}{x^3} \ln{(x)} \text{ H}(x-1)~,
    \label{eq:approx_pair_annihilation}
\end{equation}
where $H(x-1)$ is the Heaviside function. The approximation reproduces accurately the behaviour near the peak at $x_\textup{peak} \sim 3.7$ and over the range $1.3 < x < 10^{4}$ which usually is the most relevant during the calculations.

\subsubsection*{Arrival times and observed frequencies}
The time when the observer receives the radiation produced at radius $R$ is computed applying equation 26 in \cite{nava13}, which has been derived under the assumption that the radiation is dominated by matter at $\cos\theta=\beta$ (in this case the Doppler factor is equal to $\Gamma$):
\begin{equation}
    t = (1+z)~\left[\int_0^{R} \frac{1-\beta_{\rm sh}}{\beta_{\rm sh}c}dr + \frac{R}{\Gamma^2(1+\beta)c}\right]~,
\end{equation}
where $\beta_{\rm sh}$ and $\beta$ are the shock and fluid velocity, respectively, in units of $c$.
This time results from the sum of a radial delay, i.e. the difference between light travel time to radius $R$ and shock expansion time to the same radius, and an angular delay for photons emitted from the same radius $R$ but at $\cos\theta=\beta$ with respect to the line of sight.

Consequently, for the blueshift of the observed frequencies we adopt a Doppler factor $\delta=\Gamma$. 

\subsubsection*{Jet geometry and jet break}
The jet is assumed to be a cone with semi-aperture angle $\theta_{\rm jet}$ and top-hat geometry. The observer is assumed to be located along the jet axis. When $1/\Gamma\sim\theta_{\rm jet}$, the flux light-curves steepens, due to geometrical effects and possibly to side-ways expansion. In our code we neglect the latter and consider only the geometrical effect. 

For GRB~201216C, we find that a jet as narrow as $\theta_{\rm jet}\sim1^\circ$ is needed to avoid overestimating the radio emission. Smaller jet opening angles (in the range 0.6-1$^\circ$) would still provide a very good fit of MAGIC, X-ray and optical observations and predict a lower flux in the radio band, where observations are available only at late times ($>5$\,days).

\subsubsection*{Computation time and data modeling}
The numerical code is computationally expensive. The time needed to compute one realization (i.e. infer spectra from radio to TeV energies and from a few seconds to several days) goes from about ten seconds to a few hours, depending mostly on the strength of the magnetic field (and hence on density and $\epsilon_{\rm B}$): short cooling times require shorter time steps in evolving the equation for particle evolution, and hence longer computation time.

To find a good modeling to the data, we thus start from a coarse grid of values for the parameters $E_{\rm k}, A_\star$ (or $n_0$), $\epsilon_{\rm e}$, $\epsilon_{\rm B}$ and $p$. The ranges investigated are reported in Table~\ref{tab:model_parameters}.
After eliminating parts of the parameter space that give predictions completely inconsistent with observations, we repeated the simulations on narrower ranges of values and with a finer grid, until we found a good description of  the data. 
We note that the values of $\Gamma_0$ and $\theta_{\rm jet}$ affect only the initial (before deceleration) and final part of the light-curve. To save computation time, in our initial search of good solutions they are kept fixed to reasonable values and their values are adjusted only after good solutions are found, to improve the description of early time data (before $\sim300$\,seconds in the case of GRB~201216C) and of late-time data (i.e. to predict fluxes below the radio data, in case of GRB~201216).  

\newpage


\section*{Affiliations}
$^{1}$ {Japanese MAGIC Group: Institute for Cosmic Ray Research (ICRR), The University of Tokyo, Kashiwa, 277-8582 Chiba, Japan} \\
$^{2}$ {Instituto de Astrof\'isica de Canarias and Dpto. de  Astrof\'isica, Universidad de La Laguna, E-38200, La Laguna, Tenerife, Spain} \\
$^{3}$ {Instituto de Astrof\'isica de Andaluc\'ia-CSIC, Glorieta de la Astronom\'ia s/n, 18008, Granada, Spain} \\
$^{4}$ {National Institute for Astrophysics (INAF), I-00136 Rome, Italy} \\
$^{5}$ {Universit\`a di Udine and INFN Trieste, I-33100 Udine, Italy} \\
$^{6}$ {Max-Planck-Institut f\"ur Physik, D-80805 M\"unchen, Germany} \\
$^{7}$ {Universit\`a di Padova and INFN, I-35131 Padova, Italy} \\
$^{8}$ {Institut de F\'isica d'Altes Energies (IFAE), The Barcelona Institute of Science and Technology (BIST), E-08193 Bellaterra (Barcelona), Spain} \\
$^{9}$ {Technische Universit\"at Dortmund, D-44221 Dortmund, Germany} \\
$^{10}$ {Croatian MAGIC Group: University of Zagreb, Faculty of Electrical Engineering and Computing (FER), 10000 Zagreb, Croatia} \\
$^{11}$ {IPARCOS Institute and EMFTEL Department, Universidad Complutense de Madrid, E-28040 Madrid, Spain} \\
$^{12}$ {Centro Brasileiro de Pesquisas F\'isicas (CBPF), 22290-180 URCA, Rio de Janeiro (RJ), Brazil} \\
$^{13}$ {University of Lodz, Faculty of Physics and Applied Informatics, Department of Astrophysics, 90-236 Lodz, Poland} \\
$^{14}$ {Centro de Investigaciones Energ\'eticas, Medioambientales y Tecnol\'ogicas, E-28040 Madrid, Spain} \\
$^{15}$ {ETH Z\"urich, CH-8093 Z\"urich, Switzerland} \\
$^{16}$ {Departament de F\'isica, and CERES-IEEC, Universitat Aut\`onoma de Barcelona, E-08193 Bellaterra, Spain} \\
$^{17}$ {Universit\`a di Pisa and INFN Pisa, I-56126 Pisa, Italy} \\
$^{18}$ {Universitat de Barcelona, ICCUB, IEEC-UB, E-08028 Barcelona, Spain} \\
$^{19}$ {Department for Physics and Technology, University of Bergen, Norway} \\
$^{20}$ {INFN MAGIC Group: INFN Sezione di Catania and Dipartimento di Fisica e Astronomia, University of Catania, I-95123 Catania, Italy} \\
$^{21}$ {INFN MAGIC Group: INFN Sezione di Torino and Universit\`a degli Studi di Torino, I-10125 Torino, Italy} \\
$^{22}$ {INFN MAGIC Group: INFN Sezione di Bari and Dipartimento Interateneo di Fisica dell'Universit\`a e del Politecnico di Bari, I-70125 Bari, Italy} \\
$^{23}$ {Croatian MAGIC Group: University of Rijeka, Faculty of Physics, 51000 Rijeka, Croatia} \\
$^{24}$ {Universit\"at W\"urzburg, D-97074 W\"urzburg, Germany} \\
$^{25}$ {University of Geneva, Chemin d'Ecogia 16, CH-1290 Versoix, Switzerland} \\
$^{26}$ {Japanese MAGIC Group: Physics Program, Graduate School of Advanced Science and Engineering, Hiroshima University, 739-8526 Hiroshima, Japan} \\
$^{27}$ {Deutsches Elektronen-Synchrotron (DESY), D-15738 Zeuthen, Germany} \\
$^{28}$ {Armenian MAGIC Group: ICRANet-Armenia, 0019 Yerevan, Armenia} \\
$^{29}$ {Croatian MAGIC Group: University of Split, Faculty of Electrical Engineering, Mechanical Engineering and Naval Architecture (FESB), 21000 Split, Croatia} \\
$^{30}$ {Croatian MAGIC Group: Josip Juraj Strossmayer University of Osijek, Department of Physics, 31000 Osijek, Croatia} \\
$^{31}$ {Finnish MAGIC Group: Finnish Centre for Astronomy with ESO, University of Turku, FI-20014 Turku, Finland} \\
$^{32}$ {Japanese MAGIC Group: Department of Physics, Tokai University, Hiratsuka, 259-1292 Kanagawa, Japan} \\
$^{33}$ {Universit\`a di Siena and INFN Pisa, I-53100 Siena, Italy} \\
$^{34}$ {Saha Institute of Nuclear Physics, A CI of Homi Bhabha National Institute, Kolkata 700064, West Bengal, India} \\
$^{35}$ {Inst. for Nucl. Research and Nucl. Energy, Bulgarian Academy of Sciences, BG-1784 Sofia, Bulgaria} \\
$^{36}$ {Finnish MAGIC Group: Space Physics and Astronomy Research Unit, University of Oulu, FI-90014 Oulu, Finland} \\
$^{37}$ {Japanese MAGIC Group: Chiba University, ICEHAP, 263-8522 Chiba, Japan} \\
$^{38}$ {Japanese MAGIC Group: Institute for Space-Earth Environmental Research and Kobayashi-Maskawa Institute for the Origin of Particles and the Universe, Nagoya University, 464-6801 Nagoya, Japan} \\
$^{39}$ {Japanese MAGIC Group: Department of Physics, Kyoto University, 606-8502 Kyoto, Japan} \\
$^{40}$ {INFN MAGIC Group: INFN Sezione di Perugia, I-06123 Perugia, Italy} \\
$^{41}$ {INFN MAGIC Group: INFN Roma Tor Vergata, I-00133 Roma, Italy} \\
$^{42}$ {Japanese MAGIC Group: Department of Physics, Konan University, Kobe, Hyogo 658-8501, Japan} \\
$^{43}$ {also at International Center for Relativistic Astrophysics (ICRA), Rome, Italy} \\
$^{44}$ {now at Institute for Astro- and Particle Physics, University of Innsbruck, A-6020 Innsbruck, Austria} \\
$^{45}$ {also at Port d'Informaci\'o Cient\'ifica (PIC), E-08193 Bellaterra (Barcelona), Spain} \\
$^{46}$ {also at Institute for Astro- and Particle Physics, University of Innsbruck, A-6020 Innsbruck, Austria} \\
$^{47}$ {also at Department of Physics, University of Oslo, Norway} \\
$^{48}$ {also at Dipartimento di Fisica, Universit\`a di Trieste, I-34127 Trieste, Italy} \\
$^{49}$ {Max-Planck-Institut f\"ur Physik, D-80805 M\"unchen, Germany} \\
$^{50}$ {also at INAF Padova} \\
$^{51}$ {Japanese MAGIC Group: Institute for Cosmic Ray Research (ICRR), The University of Tokyo, Kashiwa, 277-8582 Chiba, Japan} \\
$^{52}$ {Center for Astrophysics and Cosmology, University of Nova Gorica, Vipavska 11c, 5270 Ajdov\v{s}\v{c}ina, Slovenia} \\
$^{53}$ {Department of Physics, University of Bath, Claverton Down, Bath, BA2 7AY, UK} \\
$^{54}$ {INAF-Osservatorio Astronomico di Brera, Via E. Bianchi 46, I-23807 Merate (LC), Italy} \\
$^{55}$ {European Space Agency, European Space Astronomy Centre, 28692 Villanueva de la Can\~{n}ada, Madrid, Spain}\\
$^{56}$ {Astrophysics Research Institute, Liverpool John Moores University, Liverpool Science Park IC2, 146 Brownlow Hill L3 5RF, UK}\\
$^{57}$Steward Observatory, University of Arizona, 933 North Cherry Avenue, Tucson, AZ 85721-0065, USA\\

\bsp	
\label{lastpage}
\end{document}